\documentclass{article}
\usepackage{caption}
\usepackage{graphicx}
\usepackage{dcolumn}
\usepackage{bm}

\begin{document}

\title{Economic irreversibility in pandemic control processes: 
Rigorous modeling of delayed countermeasures and consequential cost increases}



\author{Tsuyoshi Hondou$^{1*}$\\
$^{1}$Graduate School of Science, Tohoku University, Sendai 980-8578, Japan
}

\date{}
\maketitle
\noindent
$^{*}$ Corresponding author \\
E-mail: hondou@mail.sci.tohoku.ac.jp\\
\vspace{0.5cm}

\noindent
 
\begin{abstract}
After the first lockdown in response to the COVID-19 outbreak, many countries faced difficulties in balancing infection control with economics. Due to limited prior knowledge, economists began researching this issue using cost-benefit analysis and found that infection control processes significantly affect economic efficiency. A UK study used economic parameters to numerically demonstrate an optimal balance in the process, including keeping the infected population stationary. However, universally applicable knowledge, which is indispensable for the guiding principles of infection control, has not yet been clearly developed because of the methodological limitations of simulation studies. 
Here, we propose a simple model and theoretically prove the universal result of economic irreversibility by applying the idea of thermodynamics to pandemic control. This means that delaying infection control measures is more expensive than implementing infection control measures early while keeping infected populations stationary. This implies that once the infected population increases, society cannot return to its previous state without extra expenditures. This universal result is analytically obtained by focusing on the infection-spreading phase of pandemics, and is applicable not just to COVID-19, regardless of ``herd immunity.'' 
It also confirms the numerical observation of stationary infected populations in its optimally efficient process. Our findings suggest that economic irreversibility is a guiding principle for balancing infection control with economic effects. 
\end{abstract}


\maketitle

\clearpage

\section*{INTRODUCTION}
Governments in several countries fear adverse economic effects and have hesitated to take measures to control the COVID-19 infection because the economic effects may result in illness and death in the non-infected population \cite{R}.
Several economists, perceiving a serious lack of knowledge about the relationship between infection control and the economy \cite{B&D}, 
 started studying this issue from spring 2020 \cite{CEPR,R&M,ERT,Acemoglu}. 
 Rowthorn \cite{R}, along with his colleague Maciejowski \cite{R&M}, 
 utilized cost-benefit analysis (CBA) \cite{Boardman,Nas} to determine how infection control intervention costs could efficiently be utilized for inhibition of infection.
 Using the susceptible-infected-recovered (SIR) model to simulate the epidemic \cite{1}, they discussed several infection control processes to determine the optimal process. 
 The optimal process includes the stationary state of the constant infected population in its principal part. These results were obtained using numerical simulation because Rowthorn \cite{R} assumed that an explicit solution was unavailable for this issue.

 While the methodology and results of this study \cite{R,R&M}
are pioneering and significant, they are not straightforward enough to generalize because the study investigated specific situations with given parameter sets. 
 Therefore, explicit solutions independent of specific parameters are needed to reveal their universal property.
 Explicit solutions could be applicable in the United Kingdom and other countries during different situations, including the COVID-19 and other pandemics. From a physics perspective, optimization in CBA is similar to finding the minimum state of energy.
 In addition, the finding \cite{R,R&M} that the most efficient processes include the stationary state suggests a structure analogous to thermodynamic irreversibility.

 In this study, we analytically show the basic property of economic cost in infection control processes by analyzing the cyclic processes of infection control in a simple model. This model assumes 1) intensity-dependent infection control cost, and 2) exponential growth of the infected population. Although it excludes more realistic effects that may modify its results, such as the effects of special inhomogeneity of infection distribution and influx of infected persons from outside the targeted area, the simple model clearly shows the fundamental property commonly underlying diverse pandemic control processes. 
 For this purpose, we restrict ourselves to the infection-spreading phase in the pandemic model, in which the infected population grows exponentially in the absence of infection control. 
 In several pandemics, including COVID-19, the society may not arrive at a traditional immune state called ``herd immunity,'' as indicated by some studies \cite{To,Tillett}. 
 However, the infection-spreading phase is universal and principal, irrespective of whether herd immunity exists.
 Thus, the following results do not depend on the specific pandemic model. 
 By comparing the stationary state of a constant infected population, we derive several explicit solutions and inequalities of costs in infection control processes and show economic irreversibility in infection control.
 With these explicit results, we demonstrate that delaying infection control measures is more expensive than implementing early measures while keeping the infected population stationary. We will discuss the robustness of the result in the final section.

\section*{Methods}

\subsection*{Formulation with CBA}
 
 Infection control comprises measures taken to decrease the number of people infected by an individual. The average number within society is called the ``effective reproduction number," $R_t$ \cite{BF}. 
 When $R_t$ drops below 1, epidemics subside.
 Several measures, including handwashing, wearing of masks, suspension of business activities, and lockdowns can be taken to reduce $R_t$ from its uncontrolled (natural) value, $R_N (>1) $. $R_N $ equals the basic reproduction number $R_0$ \cite{BF} for the initial phase of infection.
 These measures have a negative influence on the economy and society \cite{R}. The social cost, $\hat{C}$, is positively correlated to the strength of the measures. 
 Rowthorn \cite{R} assumed that the infection control measure is taken through the value of $q$ as $R_t = R_N (1-q)$, where $q$ represents the intensity of social intervention against pandemics.
Then, he defined the social cost per unit of time as a function of $q$: $\hat{C}=\hat{C}(q)$ \cite{R,R&M}. He assumed $\hat{C}(0)=0$ because there is no infection control at $q=0$.
 
 Here, we consider the social cost induced by the infection measures as a function of the effective reproduction number $R_t$ instead of $q$. 
While Rowthorn \cite{R} assumes the maximum strength $q_{\mbox{\footnotesize{max}}}$, which corresponds to the minimum effective reproduction number $R_t$, 
we do not adopt this inessential assumption.
Our functional form of $C(R_t)$ itself is different from $\hat{C}(q)$, while the basic assumptions in Eqs.(1)--(4) are essentially the same as in Rowthorn \cite{R}.
Hereafter, we refer to the social cost per unit time as ``intervention cost'' in the form of $C(R_t)$. The following are assumed in the function $C(R_t)$.

 The condition without intervention measures corresponds to $R_t=R_N$, 
 in which $C(R_N) = 0$.
The cost should increase as the effective reproduction number decreases. 
The rate of increase in $C(R_t)$ should also increase as the effective reproduction number decreases. 
This is because society can take cost-effective measures, such as handwashing, to achieve a small decrease in $R_t$. 
If society must further decrease $R_t$, it must take costlier measures \cite{R}. 
Thus, we can set the following conditions on the intervention cost function $C(R_t)$ ($ 0 < R_t \le R_N$), where an example is shown in Fig. 1.
\begin{equation}
 C(R_t) \mbox{ is twice continuously differentiable},
\end{equation}

\begin{figure}
\includegraphics[scale=0.7]{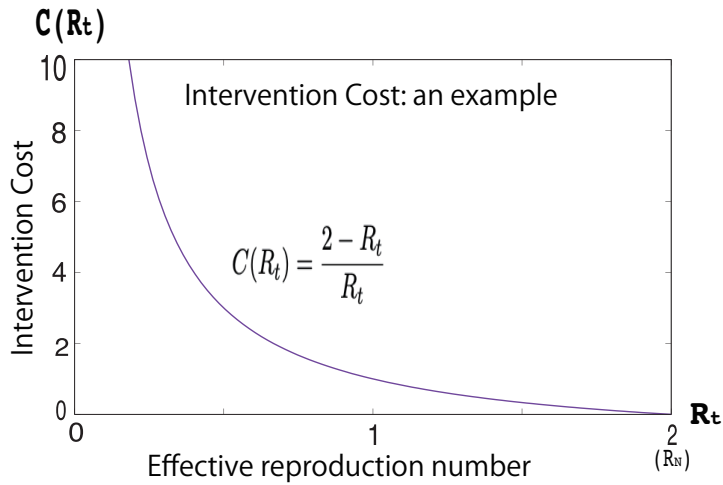}
\captionsetup{labelformat=empty,labelsep=none}
\caption{{\bf Fig. 1: An example of intervention cost $C$.}
{\small Here $R_N = 2$.}
}
\end{figure}

\begin{equation}
C(R_N) =0, 
\label{naturalcost}
\end{equation}

\begin{equation}
\frac{d C(R_t)}{d R_t} \le 0 ,
\end{equation}

\begin{equation}
\frac{d^2 C(R_t)}{d R_t^2} \ge 0 .
\label{ddc}
\end{equation}

The measure taken by spending the intervention cost $C(R)$ is to decrease the infected population $I$ (number of infected persons who are capable of transmitting infections).
The more the infected population decreases for fixed intervention costs, the more society benefits from the measure. 
The ``benefit of a decrease in the infected population'' is evaluated as the ``decrease in the cost of the infected population.''
We set this ``infection cost'' $M$ to be proportional to the infected population $I$. It includes medical costs and losses incurred by infected patients. This yields 
\begin{equation}
M(t) = c_1 I(t), 
\label{medical}
\end{equation}
where $c_1$ is a constant. 
This assumption is also the same as in Rowthorn \cite{R}.
The total cost per unit of time is the sum of intervention costs and infection costs, that is, $C(t) + M(t)$.
The optimization issue is to find $R(t)$ that minimizes the integrated total cost over a certain period,
\begin{equation}
\int [C(t) + M(t)] dt.
\end{equation}
This is equivalent to finding $R(t)$ that minimizes the average of the total cost
$\langle C(t) + M(t) \rangle$
over a certain period.

 To find the optimized intervention process specified by a protocol of $R(t)$ for a targeted period, we must consider the dynamics of the infected population. 
 Here, we begin with the SIR model proposed by Kermack and McKendrick \cite{1} because most previous studies, including Routhorn et al. \cite{R,R&M}, assumed that it is the simplest fundamental model that describes the basic dynamics of epidemics. 
 It models the exponential growth of the infected population in the outbreak stage, the peak of the infected population, and transition to the end stage \cite{Martcheva}. However, it should be noted that the following results are not restricted to the SIR framework but are expected to be generic for pandemics, as will be described later.

 \subsection*{DYNAMICS OF PANDEMICS}
 We start with the SIR model for pandemic dynamics considering its simplicity and popularity.
 The model comprises a set of differential equations that describes the epidemic disease propagation, in which the population is divided into three states:
 $S(t)$, the population ratio of susceptible persons; $I(t)$, the ratio of infected persons; and $\hat{R}_{\mbox{rec}}(t)$, the ratio of those who have recovered (or died).
 This formulation considers a closed population that is conserved. 
Note that we use the notation $\hat{R}_{\mbox{rec}}$ for recovered persons, instead of the conventional notation $R$, because we use $R_t$ for effective reproduction number.
 
\begin{equation}
\frac{d S(t)}{d t} = - \beta S(t) I(t) ,
\label{S}
\end{equation}

\begin{equation}
\frac{d I(t)}{d t} = \beta S(t) I(t) - \gamma I(t), \\
\label{I}
\end{equation}

\begin{equation}
\frac{d \hat{R}_{\mbox{rec}}(t)}{d t} = \gamma I(t),
\label{R}
\end{equation}
where $\beta$ and $\gamma$ are the infection and recovery rates, respectively.
The sum of the three population ratios remains constant: 
\begin{equation}
 S(t) + I(t) + \hat{R}_{\mbox{rec}}(t) = 1 .
\label{sum}
\end{equation}
Because of this conservation law, the model includes two independent variables.

In the following, we evaluate the infected population $I(t)$.
Eq. (\ref{I}) leads to
\begin{equation}
\frac{d I (t) }{d t} = \gamma \left[\frac{\beta S(t)}{\gamma} -1 \right] I(t) .
\label{dIdt}
\end{equation}
We restrict ourselves to the period before, but in the vicinity of, the infection peak, because this period is the most important and universal characteristic of pandemics, as will be discussed later. In this period, $S(t)$ is replaced by $S(0)$. 
This approximation is accurate in major parts of the first outbreak and its recurrent phases \cite{Vynnycky}, as shown in Fig. S1 in S1 Appendix.
Because of this approximation, the number of independent variables in this model is reduced to one.
Then, Eq.~(\ref{dIdt}) leads to
\begin{equation}
\frac{d I (t) }{d t} = \gamma \left[ \frac{\beta S(0)}{\gamma} -1\right] I(t) .
\label{first}
\end{equation}
We restrict ourselves to a fixed $\gamma$ as in Rowthorn \cite{R}.
If the set of parameters $\frac{\beta S(0)}{\gamma} > 1 $, the infections start spreading in Eq.~(\ref{first}) \cite{Brauer}. The change in the infection rate $\beta$ in $\frac{\beta S(0)}{\gamma}$ changes the dynamics of the pandemic.
The set of parameters is the effective reproduction number:
\begin{equation}
 R_t = \frac{\beta S(0)}{\gamma},
\end{equation}
where $R_t$ corresponds to the basic reproduction number $R_0$ if the following two assumptions are satisfied: 1) $\beta$ has an uncontrolled value and 2) $S(0) =1$.
The infected population increases when $R_t > 1$ and decreases for $R_t < 1$.

With $\Delta_R = R_t -1 $, Eq. \ref{first} becomes 
\begin{equation}
\frac{d I (t) }{d t} = \gamma \Delta_R \, I(t) .
\label{dd}
\end{equation}
 At $R_t=1$, the infected population is stationary as $\Delta_R = 0$.
The infection-spreading phase of pandemics generally obeys exponential dynamics \cite{Vynnycky}, characterized by the reproduction number, except in the vicinity of the infection peak. Thus, the following results are not restricted to a specific model but apply to the entire system of exponential dynamics (see S1 Appendix).
 In this formulation,
 the infected population $I(t)$ is the only variable that describes the state of the system.
 In the following sections, we will show the universal properties of the system of exponential dynamics by analyzing the cyclic process of the state variable $I(t)$.

\section*{Results}
\subsection*{IRREVERSIBLE COST IN ON/OFF-TYPE INTERVENTION PROCESSES}
Let us start the analyses of pandemic control processes.
First, we evaluate the costs of on-/off-type infection control
(see Fig. 2) 
and compare them with the costs of keeping the infected population stationary, where we assume that both processes have the same average effective reproduction number $\langle R_t \rangle =1$. Similar to thermodynamic irreversibility, comparison of the stationary and non-stationary processes will show how the pandemic control processes affect economic irreversibility.
The present on-/off-type intervention forms a cycle of both $R_t$ and $I(t)$, as shown below, where a set of lockdown and outbreak recurrences is an extreme example. 

\begin{figure}[h]
\includegraphics[scale=0.4]{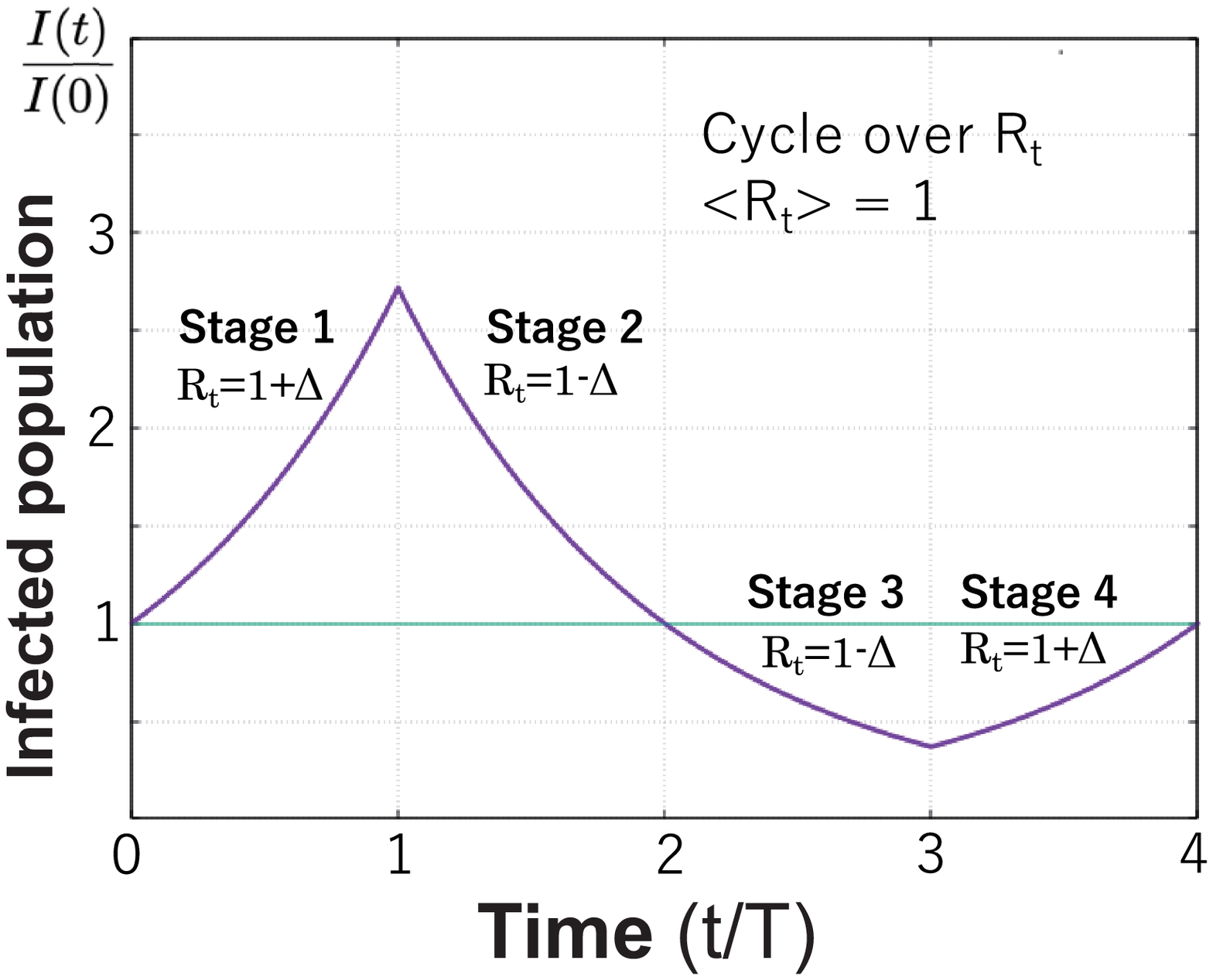}
\captionsetup{labelformat=empty,labelsep=none}
\caption{{\bf Fig. 2.
Tracing infected population during the cyclic process of infection control.} \\
{\small It is shown that the infected population is also cyclic and returns to the initial state at the end of the cycle.
The average infected population $\langle I(t) \rangle$ over the cycle is larger than that for keeping the infected population stationary. Here, we use $\gamma \Delta = 1$ in Eq.(\ref{T}).}
}
\end{figure}

 We set the amplitude of the cycle in the effective reproduction number at around $R_t = 1$ as ``$\Delta$,'' where $\Delta = |R_t - 1|$.
The cyclic process (with time interval $T$) is as follows:
\begin{enumerate}
\item[]
\mbox{Stage 1)} \, $0 < t < T$:
$I_0$ $\rightarrow$ $I_1 ( > I_0)$ with $R_t=1+\Delta$,
\item[]
\mbox{Stage 2)} \, $T < t < 2T$:
$I_1$ $\rightarrow$ $I_0$ with $R_t=1-\Delta$,
\item[]
\mbox{Stage 3}) \, $2T < t < 3T$:
$I_0$ $\rightarrow$ $ I_3 ( < I_0)$ with $R_t=1-\Delta$,
\item[]
\mbox{Stage 4)} \, $3T< t < 4T$:
$I_3$ $\rightarrow$ $I_0$ with $R_t=1+\Delta$.
\end{enumerate} 


By integrating Eq.~(\ref{dd}) from $t=0$ to $T$ with $R_t = 1 + \Delta$, we obtain the infected population $I$ at the end of Stage 1:
\begin{equation}
I(T) = I_0 e^{\gamma T \Delta}.
\label{T}
\end{equation}
Similarly, replacing $\Delta_R$ in Eq.~(\ref{dd}) by ``$- \Delta$'' and using Eq.~(\ref{T}), we obtain $I(2T)$ at the end of Stage 2:

\begin{equation}
I(2T) = I_0.
\end{equation}
Stages 3 and 4 also yield
\begin{equation}
I(4T) = I_0.
\end{equation}
We have confirmed that Stages 1 through 4 form a typical cyclic process of the state variable $I(t)$ around a stationary state kept by $R_t = 1$,
where the infected population returns to its original value. 

We calculate the average infected population to evaluate the infection cost in the cycle.
Using Eqs.~(\ref{dd}) and (\ref{T}), we have, for Stages 1 and 2,
\begin{equation}
\int_0 ^T I_{\mbox{\scriptsize Stage1}}(t) dt + \int_T ^{2T} I_{\mbox{\scriptsize Stage2}}(t) dt 
= I_0 \left[\int_0 ^T e^{\gamma \Delta \, t} dt + \int_T ^{2T} e^{\gamma \Delta \, T} e^{- \gamma \Delta \, (t-T)} dt \right]
= I_0 \int_0 ^T [e^{\gamma \Delta \, t} + e^{\gamma \Delta \,(T- t)} ] dt . 
\end{equation}

Similarly, for Stages 3 and 4, we have
\begin{equation}
\int_{2T} ^{3T} I_{\mbox{\scriptsize{Stage3}}} (t) dt + \int_{3T} ^{4T} I_{\mbox{\scriptsize Stage4}} (t) dt 
= I_0 \int_0 ^T [e^{-\gamma \Delta \, t} + e^{\gamma \Delta \,(t- T)} ] dt .
\end{equation}
Thus, we obtain the average infected population 
\begin{equation}
\frac{1}{4T}\int_0 ^{4T} I(t) dt 
=
 \frac{ I_0}{\gamma \Delta T} \sinh{(\gamma \Delta T)} 
= I_0 +\frac{ I_0 (\gamma \Delta \, T)^2}{3! } + O ((\gamma \Delta \, T)^4) \, .
\end{equation}

The stationary infected population at $R_t=1$ during the same period $4T$ is $I_0$.
This proves that the average infected population in this cycle is always higher than that of the stationary state.
Through Eq. (\ref{medical}), this result directly yields:
\begin{equation}
\langle M \rangle_{\mbox{\footnotesize{cycle}}} \, > \, \langle M \rangle_{R_t=1} ,
\label{medicalcost}
\end{equation}
where $\langle M \rangle$ denotes the time-average of the infection cost $M$.
Thus, the average infection cost for this cycle is higher than that of the stationary state.
Fig. 3 shows how the average infection cost depends on the amplitude of the cycle $\Delta$.

\begin{figure}[h]
\includegraphics[scale=0.6]{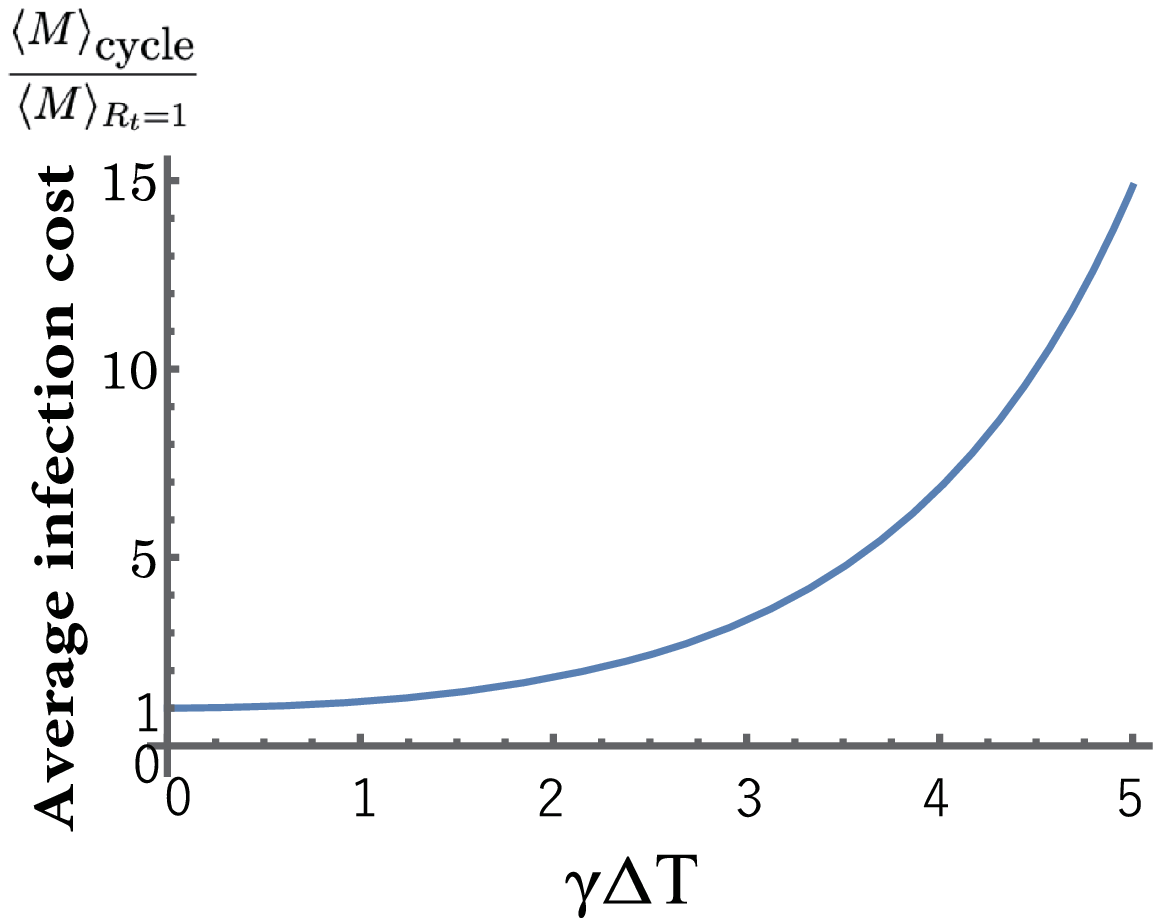}
\captionsetup{labelformat=empty,labelsep=none}
\caption{{\bf Fig.3. 
Large oscillation of intervention results in large infection cost.}
\\
{\small The average infection cost, $\langle M(I(t)) \rangle$, increases monotonically and exponentially as the amplitude of $R_t$ in the cycle $\Delta$ increases. 
The vertical axis is normalized by the average infection cost for the stationary state with $R_t = 1$, having an average effective reproduction number equal to that of the cycle.
As the state variable $I(t)$ returns to its initial state in the cycle, the increase in average infection cost is irreversible. }}
\end{figure}

Next, we calculate the average intervention cost during the cycle. 
The average intervention cost, weighing the two effective reproduction numbers $R_t =1+ \Delta$ and $R_t=1 - \Delta$ equally ($\Delta > 0)$ for the same period is
\begin{equation}
\langle C(R_t) \rangle_{\mbox{\footnotesize{cycle}}} = \frac{C(1 + \Delta) +C(1 - \Delta)}{2}.
\label{formula}
\end{equation}
The cost $ C(1 + \Delta) $ is evaluated as follows:
\begin{equation}
C(1 + \Delta) = C(1) + \int_1 ^{1+\Delta} \frac{d C(R_t)}{d R_t} dR_t \, .
\end{equation} 
From Eq. (\ref{ddc}), we find
\begin{equation}
\frac{d C(R_t)}{d R_t} > \left.\frac{d C(R_t)}{d R_t}\right|_{R_t=1} \mbox{(for $ 1 < R_t \le R_N $)} .
\end{equation}
Then, we have
\begin{equation}
C(1 + \Delta) > C(1) + \left.\frac{d C(R_t)}{d R_t}\right|_{R_t=1} \, \Delta .
\label{Delta1}
\end{equation}
Since $\left.\frac{d C(R_t)}{d R_t} < \frac{d C(R_t)}{d R_t}\right|_{R_t=1}$ for 
$0 < R_t < 1$,
\begin{equation}
C(1- \Delta) > C(1) - \left.\frac{d C(R_t)}{d R_t}\right|_{R_t=1} \, \Delta .
\label{delta2}
\end{equation} 

We obtain through Eqs. (\ref{Delta1}) and (\ref{delta2}) that
\begin{equation}
\langle C(R_t) \rangle_{\mbox{\footnotesize{cycle}}} = \frac{C(1 + \Delta) +C(1-\Delta)}{2} > C(1),
\label{for}
\end{equation}
in which $C(1)$ equals the intervention cost in a stationary state with $R_t = 1$.
Thus, we find that the average intervention cost $\langle C(R_t) \rangle$ is also higher in this cycle than in a stationary state with $R_t=1$.
Fig. 4 illustrates how the intervention cost depends on the amplitude of the cycle $\Delta$, where we use the model in Fig. 1.

\begin{figure}[h]
\includegraphics[scale=0.6]{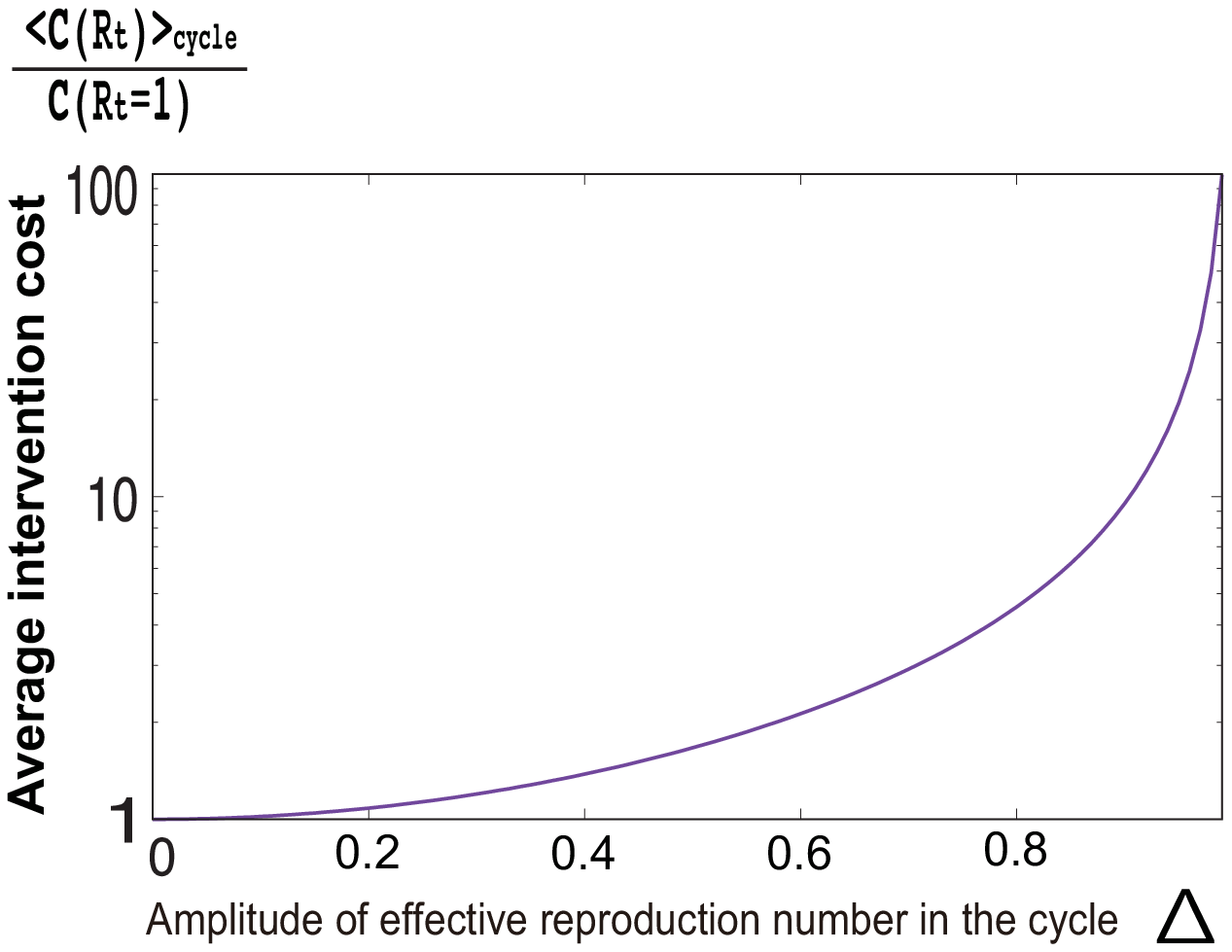}
\captionsetup{labelformat=empty,labelsep=none}
\caption{{\bf Fig. 4.
Large oscillation of intervention also results in a large intervention cost.}
\\
{\small The average intervention cost $\langle C(R_t) \rangle$ increases exponentially as the amplitude of $R_t$ in the cycle $\Delta$ increases. 
The vertical axis is normalized by the average intervention cost for the stationary state with $R_t = 1$, having an average effective reproduction number equal to that of the cycle.
We use $R_N =2$ and $C(R_t)$ of Fig. 1.
The increase in average intervention cost in the cycle does not contribute to the benefit (decrease in average infection cost) at all, as Fig. 3 shows. }
}
\end{figure}

The results show that the cycle of infection control around the stationary state provokes a higher average infected population $\langle I(t) \rangle$, 
and a higher intervention cost, compared with the stationary state.
Because the variable of the state $I(t)$ finally returns to the initial state in the cycle, the cycle above results in a higher waste of social resources (intervention cost) than in a stationary state.
The economic irreversibility in which society cannot retrieve the dissipated social resource is similar to entropy production (or free energy decreases) in thermodynamics \cite{Nicolis}. 

The total cost $C(R_t) + M(t)$ for the cycle thus satisfies the inequality
\begin{equation}
\mbox{Average of the total cost of the cyclic process} > \mbox{That of the stationary process},
\label{total}
\end{equation}
even if the two processes have the same average effective reproduction number $\langle R_t \rangle = 1$.
This means that keeping the infected population stationary is better than the cyclic pandemic control process.

We have learned that society cannot produce extra benefits (decrease in infected population) in the cyclic process compared with keeping the infected population constant while it pays extra intervention costs in the cycle.
In addition, society also incurs the disadvantage (increase in infected population) in the cycle.
Note that this inequality, Eq.~(\ref{total}), holds irrespective of specific parameters, which conflicts with previous studies on the economic efficiency of infection control. 
This inequality illustrates how on-/off-type infection control against pandemics costs society.

\subsection*{IRREVERSIBLE COST OF DELAYING MEASURES}

Now, we will show the implications of economic irreversibility based on the effect of delaying measures against pandemics.
We compare the two processes with the same initial and final states (infected population) $I_0$, in which only the swiftness of the pandemic control is different.
\begin{enumerate}
\item[]
Process 1) \, Do not perform infection control initially or perform small intervention at $t=0$ with $R_t = R_a$, in which $1 < R_a \le R_N$, until some critical time ($t=t_a$) just before serious problems such as the crash of medical capacity arise.
Then, infection control is performed at $t=t_a$ to achieve a constant $R_t < 1$ to decrease $I(t)$ back to $I_0$. 
This process is similar to the combined process of Stages 1 and 2 in Fig. 2. However, the choice of $R(t)$ before and after $t=t_a$ is arbitrary.
\item[]
Process 2) \, Perform infection control to achieve $R_t=1$ immediately at $t=0$.
\end{enumerate} 
Here, we assume $R_N > 1$ for both processes.

The advantage of Process 1 is the zero or small intervention cost $C(R_a) < C(1)$ between $t=0$ and $t=t_a$. When compared with taking immediate measures, $R_t=1$ (Process 2), 
this process saves on intervention costs between $t=0$ and $t_a$:
\begin{equation}
\int_0^{t_a} [C(1)-C(R_a) ] dt .
\label{b}
\end{equation}
Thus, what matters is whether the saving on intervention costs (Eq.~(\ref{b}) at $t=t_a$) remains positive even at the final stage, $t=t_a + t_b$, when the society returns to its initial state, $I_0$.
Thus, we calculate the average intervention cost of Process 1, $\langle C (R_t) \rangle_{\mbox{\footnotesize{delay}}}$, during the period from $t=0$ to $t=t_a + t_b$.
From Eq.~(\ref{dd}), the state of $I(t)$ at $t=t_a$ is $I(t_a) = I_0 e^{\gamma t_a \Delta_a}$, where $\Delta_a = R_a - 1 $.
We assume that $I(t)$ returns to $I_0$ at $t=t_a+t_b$, and $R_t=R_b = 1 - \Delta_b \, (0 < \Delta_b < 1)$ for $t_a < t \le t_a + t_b$.
Then, we have
$I(t_a+t_b) = I(t_a) e^{-\gamma t_b \Delta_b}$.
As $I(t_a+t_b) = I_0$, we obtained the equality
\begin{equation} 
t_a \Delta_a = t_b \Delta_b.
\label{para}
\end{equation}
Then, the average intervention cost between $t=0$ and $t = t_a + t_b$ is written as
\begin{equation}
\langle C(R_t) \rangle_{\mbox{\footnotesize{delay}}} =
\frac{t_a}{t_a+t_b} C(1 + \Delta_a) + \frac{t_b}{t_a + t_b} C(1-\Delta_b).
\label{C}
\end{equation}

From Eqs.~(\ref{Delta1}) and (\ref{delta2}), Eq.~(\ref{C}) satisfies the following condition:
\begin{equation}
\langle C(R_t) \rangle_{\mbox{\footnotesize{delay}}} > \frac{t_a}{t_a+t_b} \left [C(1)+ \left.\frac{dC}{dR_t} \right|_{R_t=1} \Delta_a \right] + \frac{t_b}{t_a + t_b} \left[C(1) -\left.\frac{dC}{dR_t} \right|_{R_t=1} \Delta_b \right]. 
\label{cr}
\end{equation}
Using Eq.~(\ref{para}), the right-hand side of Eq.~(\ref{cr}) equals C(1). 
Thus, we obtain
\begin{equation}
\langle C(R_t) \rangle_{\mbox{\footnotesize{delay}}} > C(1) .
\label{ll}
\end{equation}
The right-hand side is the average intervention cost of Process 2. 
The average intervention cost $\langle C(R_t) \rangle_{\mbox{\footnotesize{delay}}} $ is found to be higher for delayed measures (Process 1) than in a stationary state (Process 2).
The inequality is universal because Eq.~(\ref{ll}) holds for any process with linear functions with parameters $\Delta_a$ and $\Delta_b$.
Furthermore, because any integrable function can be decomposed into a set of linear functions with arbitrary precision, Eq.~(\ref{ll}) holds for any process of integrable $R(t)$ on the condition that the variable of state $I(t)$ returns to its initial state.

Apparently, the infection cost satisfies the similar inequality condition as above: 
\begin{equation}
\langle M(I(R_t)) \rangle_{\mbox{\footnotesize{delay}}} > M(I(1)) ,
\end{equation}
as the average infected population is higher when control measures are delayed (Process 1) than when the infected population is held stationary with $R_t = 1$. 
The results show that a society delaying measures must incur more intervention and infection costs during the process until $I(t)$ returns to its original state, 
even if it temporarily saves on intervention costs.
In other words, once the infected population increases, the society cannot return to the previous lower-infection state without paying extra costs compared with a stationary state (see Fig. 5).
An increase in the infected population always results in economic irreversibility in pandemics, except in the vicinity of the infection peak.
The universal result of the model is again independent of the details of the system.

\begin{figure}[h]
\includegraphics[scale=0.7]{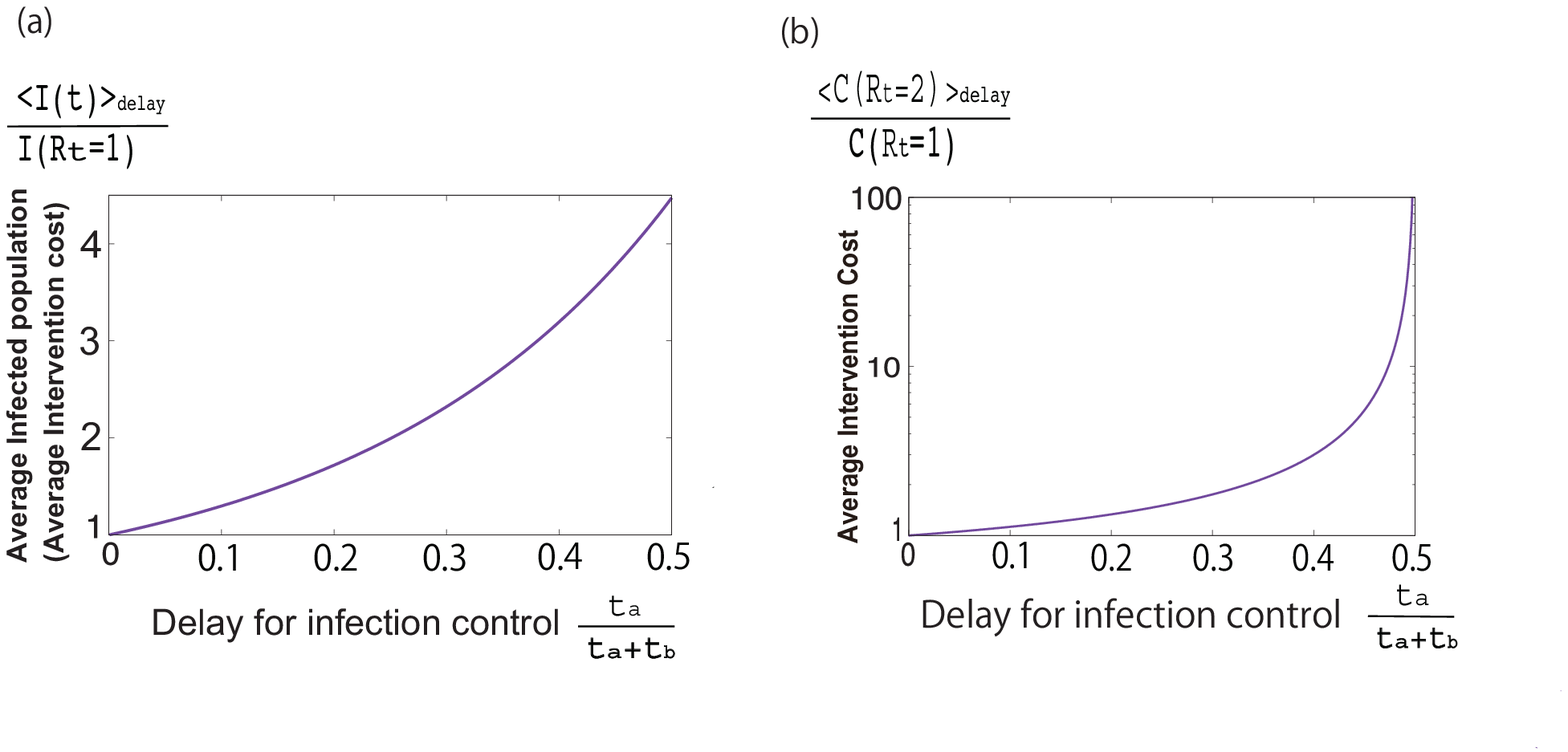}
\captionsetup{labelformat=empty,labelsep=none}
\caption{{\bf Fig. 5.
Delayed measures result in an increase in the infected population and intervention costs.}
\\
{\small The vertical axis is normalized by the (a) average infected population and (b) average intervention cost for the stationary state with $R_t = 1$.
We assume the basic reproduction number $R_N =2$ and use the model in Fig. 1 for the intervention cost $C(R_t)$, where the parameters $\gamma = \Delta_a=1$ and $t_a + t_b = 5$. The costs rapidly increase as the delay time for infection control increases. After the critical delay time, $t_a = 5/2$ in this model (corresponding to 0.5 in the horizontal axis of the figures), the system cannot return to the original infected-population state $I(t=0)$ within the period, $t_a + t_b = 5$.}
}
\end{figure}

\section*{DISCUSSION and CONCLUSION}
This study theoretically analyzed the fundamental structure of economic irreversibility in infection control processes during the infection-spreading phase.
Delaying measures against the spread of infection results in cost increases, in which sets of lockdowns and recurrences are extreme examples.
Once the state variable $I(t)$ increases, the system is irreversible because it cannot return to the previous low-infection state without extra expenditures compared with the stationary state of low infection.
These general results contradict the naive idea that infection control always results in economic damage.

The merit of keeping the infected population constant has been previously discussed by Rowthorn \cite{R}, who stated, ``The most
robust conclusion is that, if a relatively inexpensive way can be found to reduce
the net reproduction ratio to r = 1, that is, the policy to aim for in the medium term.'' His numerical finding is explained by our analytical result. 
It should be noted that our results are derived from a simple model, and so do not account for other effects such as vaccination and seasonal modulation. 
Additionally, our analysis is restricted to a principal part of the pandemic, namely, the infection-spreading phase.
These are the limitations of our study.

The validity of the present study is subject to assumptions of the methodology. 
In addition to the conventional methodological assumptions of a homogeneous mixing of the infected and susceptible populations \cite{JG} and constant rates \cite{1}, we made two principal assumptions:
\begin{enumerate}
\item[1.]
The intervention cost depends on the effective reproduction number $R_t$, and its cost function $C(R_t)$ is concave, as in Eq. (\ref{ddc}).
\item[2.]
The epidemic is in the infection-spreading phase, and thus increases and decreases in the infected population while obeying exponential dynamics, as in Eq.~(\ref{dd}).
\end{enumerate}
The first assumption is the same as that in previous research \cite{R,R&M} through the relation $R_t = R_N (1-q(t))$, which is intuitive, as shown in section II. 
The exponential dynamics in the second assumption is a common feature of pandemics, as clearly illustrated in S1 Appendix. 
This feature is intuitively understandable, as infectability in pandemics is generally characterized by the reproduction number. 
The results are not restricted to a specific model but are common features in most pandemics, as long as the infection-spreading phase is expected to last longer than the time scale of variation of the infected population.

Our study does not offer concrete cost values such as the conventional CBA.
However, our results reveal the common structure of the underlying costs in more complex/realistic models, because our simple model shares their fundamental assumptions.
The universal character found in this study is similar to thermodynamics \cite{Callen}.
The theory of thermodynamics alone does not reveal the physical quantity of a system.
However, it provides a quantitative relationship among physical variables and shows physical irreversibility. Physical irreversibility is similar to the present result that an increase in the infected population is economically irreversible.

Irreversibility of thermodynamics is caused by the deviation from thermal equilibrium. 
Carnot's cycle is known as a reversible thermodynamics process, which converts thermal energy into mechanical energy at maximum efficiency \cite{Callen}. This is analogous to the CBA in the sense that the CBA evaluates the efficiency of the conversion from social intervention cost to a benefit
 (decrease in the infected population in the present case). 
Optimal energy conversion is available in Carnot's cycle because the cycle is at equilibrium, and so there is no entropy production.
In a nonequilibrium stationary state, 
it requires a finite cost to keep the system stationary \cite{Hondou-Sekimoto,Hondou-Takagi}, in which the efficiency of energy conversion is different from that at equilibrium.
However, even if the system is out of equilibrium, the efficiency of energy conversion \cite{Kame-Hondou} and an equality on irreversible work and free energy difference \cite{Jarzynski} can be analytically discussed using the concepts and methodology of thermodynamics and statistical mechanics. 
The present system corresponds to a nonequilibrium, even in the stationary state of a constant infected population, 
because stationarity is maintained by incurring infection control costs, with $C(R_t = 1) > 0$, to inhibit an increase in the infected population.
Therefore, application of the concepts and methodology of nonequilibrium thermodynamics to the CBA would be challenging \cite{Tsirlin} because of economic irreversibility \cite{Martinas,Amir,Ayres} and its universality, as shown here.

Our analysis of the infection-spreading phase explicitly shows that the increased state is economically irreversible once the infected population increases. 
This result is applicable not just to COVID-19 and regardless of whether ``herd immunity'' exists \cite{To,Tillett}. 
To the best of our knowledge, this is the first analytical study of economic efficiency during pandemic control.
These results may provide guidance for infection control during pandemics, just as the prediction of natural phenomena and several industrial applications benefit from the principles of thermodynamics.
However, our study does not provide a solution to the question of what level the infected population should be limited to.
This question concerns whether we should aim to completely eradicate infection. Analytical studies to identify the determinants of the most effective pandemic control processes are an important challenge for the future.

\section*{Acknowledgments}
  The author wishes to acknowledge T. Onai, S. Yonemura, and K. Hirata for their fruitful discussions. S. Takagi, R. Seto, M. Arikawa, M. Sano, and H. Chat\'{e} are also acknowledged by the author for their critical reading of the manuscript and helpful comments. We would like to thank Editage for English language editing. 


\section*{Supporting information}

\section*{S1 Appendix. Robustness of exponential growth in an infection-spreading (outbreak) phase }

%
\subsection*{SIS model}
This model assumes that the infected persons do not have acquired immunity. Thus, they will again become susceptible.

\begin{equation}
\frac{d S(t)}{d t} = - \beta S(t) I(t) + \gamma I(t),
\end{equation}

\begin{equation}
\frac{d I(t)}{d t} = \beta S(t) I(t) - \gamma I(t),
\end{equation}
where $\beta$ and $\gamma$ are infection and recovery (in this case, to the susceptive state) rates, respectively.
The sum of the two population ratios remains constant: 
\begin{equation}
 S(t) + I(t) = 1.
\label{sum}
\end{equation}

\subsection*{SIRS model}
In this model, the infected persons obtain acquired immunity temporarily but become susceptible later.
\begin{equation}
\frac{d S(t)}{d t} = - \beta S(t) I(t) + h \hat{R}_{\mbox{rec}}(t),
\end{equation}

\begin{equation}
\frac{d I(t)}{d t} = \beta S(t) I(t) - \gamma I(t), \\
\end{equation}

\begin{equation}
\frac{d \hat{R}_{\mbox{rec}}(t)}{d t} = \gamma I(t) - h \hat{R}_{\mbox{rec}}(t),
\end{equation}
where $h$ is the rate of losing the temporarily acquired immunity. The sum of the three population ratios remains constant: 
\begin{equation}
 S(t) + I(t) + \hat{R}_{\mbox{rec}}(t) = 1.
\end{equation}

\subsection*{Confirmation}
In Figure S1, it is confirmed that exponential dynamics in outbreak phases are common even for different pandemic systems. 
This means that the results of this study, which assume that the dynamics are governed by the reproduction number, can be applied to any pandemic system that obeys exponential dynamics in outbreak phases.
The present results are universal in the sense that they are independent of the details of the specific pandemic.

\begin{figure}[h]
\includegraphics[scale=0.4]{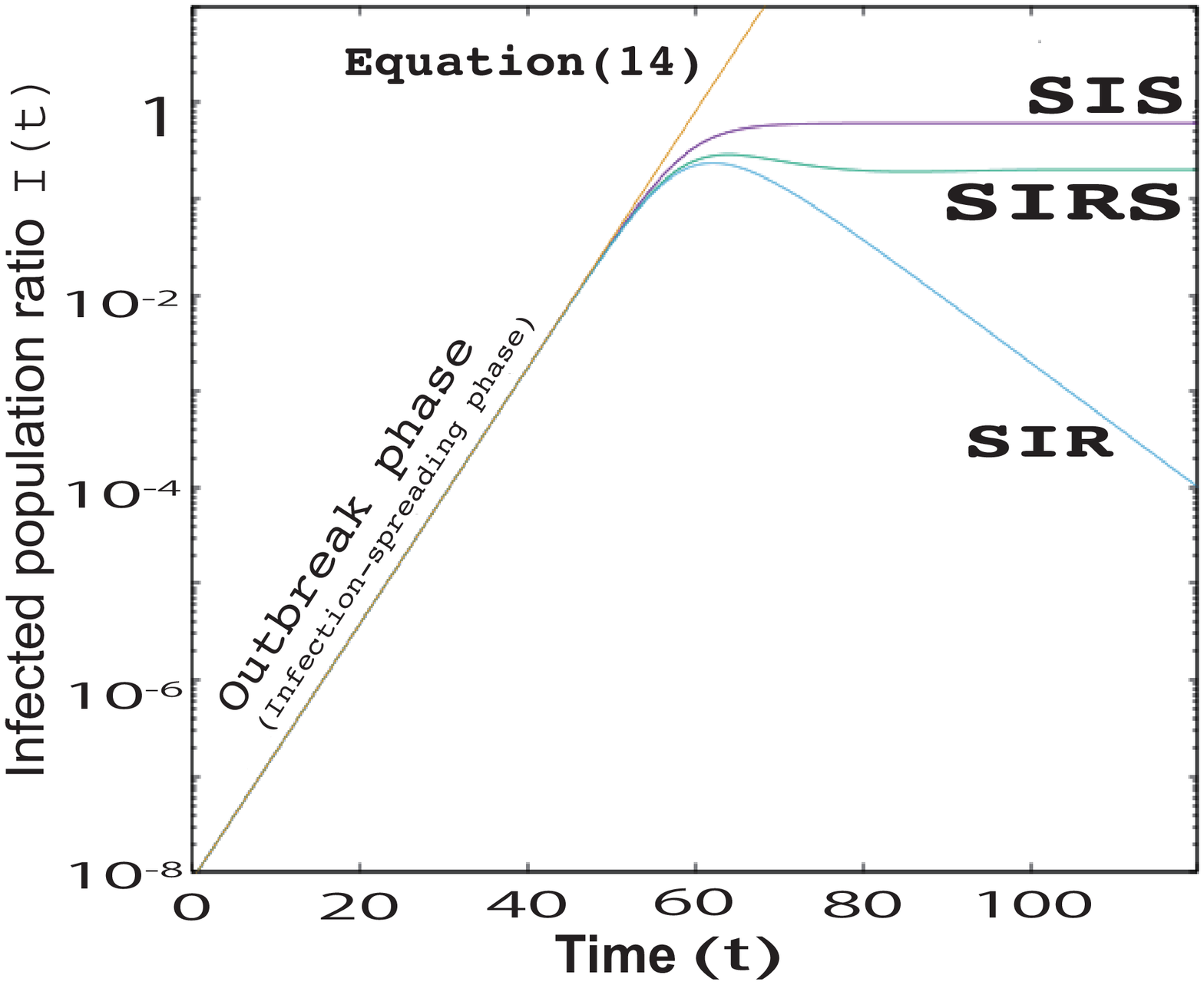}
\captionsetup{labelformat=empty,labelsep=none}
\caption{{\bf Fig S1. The dynamics of three pandemic models (susceptible-infected-recovered (SIR), susceptible-infected-susceptible (SIS), and susceptible-infected-recovered-susceptible (SIRS) models) with those of our theoretical assumptions are shown. }
\\
All four models with the same basic reproduction number are shown to be precisely the same in their outbreak phases. This is because the effective reproduction number is the only index that characterizes the pandemic dynamics of the outbreak phase.
Therefore, our methodology and results are not restricted to the specific model but are applicable to any pandemic in which the effective reproduction number characterizes the dynamics of outbreak phases. 
Here, we used $\beta = 0.51$, $\gamma = 0.204$, and $h=0.1$, which correspond to the basic reproduction number $R_0 = 2.5$. 
Numerical calculations are performed using the Euler method, in which 
the initial values are as follows: 
Total population $N = 1.2 \times 10^8 +1$, 
$S(0)=1.2 \times 10^8 /N$, $I(0) = 1/N$, $\hat{R}_{\mbox{rec}}(0)=0/N$ ($\hat{R}_{\mbox{rec}}$ is for the SIR and SIRS models only).
}
\end{figure}

\end{document}